\documentclass{article}
\usepackage[margin=1in]{geometry}

\usepackage{dsfont,multicol, booktabs,subfig, natbib}
\usepackage{amsmath,amssymb, graphics,graphicx}
\usepackage{authblk}

\newcommand{\bI}{{\mathbf I}}

\newcommand{\bX}{{\mathbf X}}

\newcommand{\bZ}{{\mathbf Z}}
\newcommand{\bgamma}{{\boldsymbol \gamma}}
\newcommand{\bPsi}{{\boldsymbol \Psi}}

\newcommand{\btheta}{{\boldsymbol \theta}}
\newcommand{\Ind}{\mathds{1}}

\title{Bayesian Distributed Lag Interaction Models to Identify Perinatal Windows of Vulnerability in Children's Health}
\date{}

\author[1]{Ander Wilson}
\author[2]{Yueh-Hsiu Mathilda Chiu}
\author[2]{Hsiao-Hsien Leon Hsu}
\author[2]{Robert O. Wright}
\author[2]{Rosalind J. Wright}
\author[3]{Brent A. Coull}
\affil[1]{Colorado State University}
\affil[2]{Icahn School of Medicine at Mount Sinai}
\affil[3]{Harvard T.H. Chan School of Public Health}

\begin{document}
\maketitle

\begin{abstract} 
Epidemiological research supports an association between maternal exposure to air pollution during pregnancy and adverse children's health outcomes. Advances in exposure assessment and statistics allow for estimation of both critical windows of vulnerability and exposure effect heterogeneity. Simultaneous estimation of windows of vulnerability and effect heterogeneity can be accomplished by fitting a distributed lag model (DLM) stratified by subgroup. However, this can provide an incomplete picture of how effects vary across subgroups because it does not allow for subgroups to have the same window but different within-window effects  or to have different windows but the same within-window effect. Because the timing of some developmental processes are common across subpopulations of infants while for others the timing differs across subgroups, both scenarios are important to consider when evaluating health risks of prenatal exposures. We propose a new approach that partitions the DLM into a constrained functional predictor that estimates windows of vulnerability and a scalar effect representing the within-window effect directly. The proposed method allows for heterogeneity in only the window, only the within-window effect, or both. In a simulation study we show that a model assuming a shared component across groups results in lower bias and mean squared error for the estimated windows and effects when that component is in fact constant across groups. We apply the proposed method to estimate windows of vulnerability in the association between prenatal exposures to fine particulate matter and each of birth weight and asthma incidence, and estimate how these associations vary by sex and maternal obesity status, in a Boston-area prospective pre-birth cohort study.\vspace{1ex}

\noindent\textbf{keywords:} Birth weight; Child asthma; Distributed lag models; Exposure effect heterogeneity; Fine particulate matter; Functional data analysis.
\end{abstract}

\section{Introduction}

A growing body of research supports an association between maternal exposure to air pollution during pregnancy and a variety of birth and children's health outcomes. Epidemiological studies have found that maternal exposure to ambient air pollution is associated with decreased birth weight as well as increased risk of preterm birth and respiratory disorders including asthma \citep{Sram2005,Kelly2011,Shah2011,Stieb2012,Savitz2014,Jedrychowski2014,Chiu2014}. The National Institute of Environmental Health Sciences (NIEHS) has identified both the estimation of windows of vulnerability and the identification of susceptible subpopulations as critical research directions in environmental health research  \citep{NIEHS2012}.  A critical methodological gap is the lack of available statistical methods to simultaneously identify  windows of vulnerability and susceptible populations.  

Windows of vulnerability are time periods during which exposure to a toxin has an increased association with current or future health status \citep{Barr2000, West2002}.  Prenatal systems development is a multi-event process progressing sequentially from early gestation \citep{Kajekar2007}. The identification of  windows of vulnerability, which in turn corresponds to sensitive stages of development, can inform our understanding of the underlying pathways through which an environmental exposure operates.  Presumably, the  window is defined by developmental specific events (i.e. gene expression changes, growth/cell density, vascularization etc.) that are transient and environmental exposure that is concurrent to these events.

The distributed lag model (DLM) framework has a long history in air pollution research and was originally developed for time-series analysis where an outcome observed on a given day is jointly regressed on exposures over a previous time period \citep{Schwartz2000a,Zanobetti2000}.  Several recent studies have applied DLMs to estimate  windows vulnerability during which air pollution has an elevated association with preterm birth \citep{Warren2012,Chang2015}, decreased birth weight \citep{Warren2013}, childhood asthma \citep{Hsu2015}, and disrupted neurodevelopment \citep{Chiu2016}.  

A critical consideration in the estimation of windows of vulnerability is that the prenatal developmental process is not homogeneous across all subgroups of individuals. For example, females display earlier fetal breathing than males \citep{Becklake1999}. Because developmental timing varies, it is natural to hypothesize that the association between prenatal exposure and health outcomes in a given subgroup may not only vary in the effect size but in the timing of the window of vulnerability. Therefore, when interest focuses on windows of vulnerability, there are at least four potential patterns of effect heterogeneity: 1) both the effect size and the timing of the window vary by subgroup; 2) only the effect size varies by subgroup; 3) only the timing of the window varies by subgroup; or 4) both the window and the effect size are the same for all subgroup. Existing methods only accommodate patterns (1) and (4), but not (2) and (3).

A DLM can estimate a common window and effect size (pattern 4) or, when stratified by group,  estimate lagged effects assuming that both the effect size and the exposure window vary by subgroup (pattern 1), \citep[e.g.][]{Hsu2015,Chiu2016}. The stratified DLM approach does not share information across groups relating to either the timing of the window or the effect size within the window. In the context of functional regression, \cite{Wei2014} proposed a functional interaction model for gene-environment interactions with a time-varying environmental exposure. This approach shares information across groups (genotype) by assuming the same  windows for each genotype but the effect is scaled by the number of major alleles (0, 1, or 2). Hence, this approach can estimate effects satisfying pattern 2 under the additional constraint that the effects are proportional to the number of major alleles. However, none of these approaches allow for the estimation across all four patterns of effect heterogeneity.

In this paper we propose a new method that provides greater flexibility in characterizing effect heterogeneity when identifying windows of vulnerability is of interest. The proposed Bayesian distributed lag interaction model (BDLIM) partitions the distributed lag function from a standard DLM  into a time component that identifies windows of vulnerability and a scale component that quantifies the magnitude of the effect within the window. The approach allows both the window and the  scale components to either vary or stay constant  across subgroups. As such, the BDLIM framework can estimate a model with any of the four effect heterogeneity patterns.  To our knowledge, models that assume the window of vulnerability is the same across subgroups but the effect within the window varies across groups, and vice versa, have not previously been considered in the literature. The proposed approach allows the user to directly answer the question of whether effect heterogeneity manifests itself via changes in the window of vulnerability, the magnitude of the effect, or both, which in turn more directly answers the question of whether an environmental exposure affects the developmental process of an infant similarly across subgroups.

In the BDLIM framework the time-varying weight function is treated as a functional predictor that is scaled by the scalar effect size. This partitioning requires that identifiability constraints be placed on the parameters of the weight function. Under certain assumptions about effect heterogeneity the model can be reparameterized to relax the identifiability constraints on the parameter space and reduces to a mixed effects model. In other cases, including the general linear model setting for discrete responses, we use a slice sampler to efficiently estimate the model from the constrained space. We make software available for BDLIM in the {\tt R} package {\tt regimes} (REGression In Multivariate Exposure Settings).

We use BDLIM to estimate the association between fine particulate matter (PM$_{2.5}$) measured weekly over pregnancy and two outcomes--birth weight for gestational age (BWGA) $z$-score and asthma incidence--in a prospective Boston-area pregnancy cohort. Following \cite{Lakshmanan2015}, we evaluate whether the association between PM$_{2.5}$ and BWGA $z$-score varies by both sex and maternal obesity status.  Following  \cite{Hsu2015}, we evaluate how effects on asthma incidence vary by sex.

\section{The ACCESS Data}\label{s:data}
We analyze data from the Asthma Coalition on Community, Environment, and Social Stress (ACCESS) project \citep{Wright2008}. ACCESS is a prospective, longitudinal study designed to investigate the effects of stress and other  environmental factors, including air pollution, on asthma risk in a urban U.S. setting. The ACCESS cohort includes data on 997 mother-child pairs that were recruited between August 2002 and January 2007.  The women were at least 18 years of age, spoke English or Spanish, and received prenatal care at one of two Boston, MA area hospitals or affiliated community health centers. To date, the ACCESS cohort has been used to study the relationship of air pollution exposures, maternal stress, and other risk factors with outcomes including asthma, wheeze, and birth weight \citep{Chiu2012, Hsu2015,Lakshmanan2015}. Like previous ACCESS studies, we limit the analysis to full-term ($\ge$37 weeks), live births with complete exposure, outcome, and covariate data.

For each child we consider BWGA $z$-scores and maternal-reported and clinician-diagnosed asthma as outcomes. The data contain maternal and child covariate information including: maternal age at enrollment; race/ethnicity (black, hispanic, and white); maternal education (two categories less than high school and high school diploma or more); self-reported smoking during pregnancy; indicator of maternal pre-pregnancy obesity; infant sex; maternal atopy (ever self-reported doctor-diagnosed asthma, eczema, or hay fever); season of birth; a previously described maternal stress index \citep{Chiu2012}; and a previously described neighborhood disadvantage index \citep{Chiu2012}. Maternal exposures of PM$_{2.5}$ were estimated with a hybrid land use regression model that incorporates satellite-derived aerosol optical depth measures \citep{Kloog2011}. Each mother was assigned an average PM$_{2.5}$ exposure value for each week of pregnancy based on the predicted value at her address of residence. We limit our analysis to exposure during the first 37 weeks of pregnancy.

\section{Bayesian Distributed Lag Interaction Models}\label{s:model}
\subsection{Approach}
Interest focuses on  estimation of the association between time-varying exposure $X_i(t)$, $t\in\mathcal{T}$, and scalar outcome $Y_i$ while controlling for a vector of baseline covariates $\bZ_i$. We denote individuals by $i=1,\dots,n$.

We  parameterize the time-varying effect of exposure as $\beta w(t)$, where $w(t)$ is a continuous weight function that captures temporal variation in the association between $X(t)$ and $Y$ and $\beta$ is the scalar effect size. The weight function $w(t)$ identifies windows of vulnerability in which the  exposure effect is elevated relative to other time periods. To allow for heterogeneity among subgroups (e.g. infant sex) indexed by $j$, we allow either or both of these quantities to vary across levels of $j$, denoted $\beta_j$ and $w_j(t)$.  Similar to a stratified DLM, this parameterization allows for group-level modification of both the window and effect size. In addition, this parameterization accommodates scenarios not yet considered whereby only the location of the window or only the magnitude of the effect, but not both, vary by group. 

\subsection{BDLIM for a single population}\label{sub:BDLIM}
In the BDLIM regression model for a single population with no effect heterogeneity we assume $E(Y_i)=\mu_i$ and 
\begin{equation}
g\left(\mu_i\right)=\alpha + \beta\int_{t\in\mathcal{T}} X_{i}(t)w(t)dt + \bZ_i^T\bgamma,
\label{eq:BDLIM}
\end{equation}
where $g(\cdot)$ is a monotone link function, $\alpha$ is the intercept, and $\bgamma$ is a vector of unknown regression coefficients for the covariates $\bZ$. The model in \eqref{eq:BDLIM} is similar to a functional linear model in which the  total effect of exposure $X(t)$ on outcome $Y$ at time $t$ is $\beta  w(t)$. We will refer to this model as BDLIM-n, where the ``n" indicates no heterogeneity between subgroups.

For identifiability, we constrain the weight function such that $\int_{t\in\mathcal{T}}  w(t)^2 dt=1$ and $\int_{t\in\mathcal{T}}  w(t)dt\ge0$. The weight function is allowed to be both positive and negative to account for exposures that are a toxicant during some time periods but a nutrient during others. The constraint $\int_{t\in\mathcal{T}}  w(t)dt\ge0$ assures that $\beta$ reflects the direction of the cumulative effect, $\beta\int_{t\in\mathcal{T}}  w(t)dt$. The constraint $\int_{t\in\mathcal{T}}  w(t)^2 dt=1$ ensures that the magnitude of $\beta$ (i.e. $|\beta|$) is identifiable.

To gain some insight into the BDLIM approach consider the case where $w(t)=1$. In this case $\int_{t\in\mathcal{T}}  X_{i}(t)w(t)dt=\bar{X}_i$ and \eqref{eq:BDLIM} becomes a linear model with scalar exposure covariate equal to the  mean exposure over the full pregnancy. However, once the weight function $w(t)$ varies with $t$, the weighted exposure $\int_{t\in\mathcal{T}}  X_{i}(t)w(t)dt$ gives greater relative weight to some time windows. These up-weighted times are considered the windows of vulnerability.

\subsection{BDLIM model for effect heterogeneity}\label{sub:heteomodel}
A key advantage of the BDLIM framework is the ability to  estimate the three hypothesized patterns of heterogeneity where either $\beta$, $w(t)$, or both $\beta$ and $w(t)$ vary by group. When either $\beta$ or $w(t)$ is constant across groups, BDLIM yields a more parsimonious model that results in more powerful tests of an interaction. 

Consider analysis of data for groups $j=1,\dots,J$. When both the effect size and the window of vulnerability are group-specific, the BDLIM model is 
\begin{equation}
g\left(\mu_i\right)= \alpha_{j_i} + \beta_{j_i}\int_{t\in\mathcal{T}}  X_{i}(t)w_{j_i}(t)dt + \bZ_{i}^T\bgamma ,
\label{eq:BDLIM-bw}
\end{equation}
where the subscript $j_i$ denotes group $j$ to which individual $i$ belongs.  We refer to this model as BDLIM-bw where ``bw'' indicates that both $\beta$ and $w(t)$ vary across groups. 

The BDLIM framework extends to the new scenarios where only  $\beta$ or $w(t)$ varies by group. If it is hypothesized that groups share a common window, e.g. first trimester, but the groups are differentially susceptible within that window, the model is
\begin{equation}
g\left(\mu_i\right)= \alpha_{j_i} + \beta_{j_i}\int_{t\in\mathcal{T}}  X_i(t)w(t)dt + \bZ_{i}^T\bgamma ,
\label{eq:BDLIM-b}
\end{equation}
which we denote BDLIM-b. Alternatively, if the effect is the same across groups but the windows are different, perhaps shifted by a few weeks, then the model, denoted BDLIM-w, is
\begin{equation}
g\left(\mu_i\right)=  \alpha_{j_i} + \beta\int_{t\in\mathcal{T}}  X_{i}(t)w_{j_i}(t)dt + \bZ_i^T\bgamma .
\label{eq:BDLIM-w}
\end{equation} 
Both BDLIM-b and BDLIM-w estimate patterns of effect heterogeneity not previously addressed.

\subsection{Parameterization of the functional components}\label{sub:paramfunc}
We assume a truncated basis function representation of both $X_i(t)$ and $w(t)$. Common choices for the basis expansions include splines, wavelets, Fourier series, and principal components (PCs). We use the first $K$ PCs of the covariance matrix of $X(t)$ as the basis to represent both $X(t)$ and $w(t)$ for all groups. Here, $K$ is chosen to be the number of PCs that explain a prespecified proportion of the total variability in the exposures, for example 99\% of the total variation. Hence, $X_i(t)=\sum_{k=1}^{K}\xi_{ik}\psi_k(t)$ and $w(t)=\sum_{k=1}^{K}\theta_k\psi_k(t)$. 

In practice, we observe the exposures measured over a discrete grid, in our case $t=1,\dots,37$ weeks of pregnancy. Let $\bX$ be a $n\times T$ matrix with row $i$ being the observed exposures for individual $i$ measured at times $t=1,\dots,T$, $\bX_i=(X_{i1},\dots,X_{iT})$. We use the covariance matrix of $\bX$ to estimate the basis functions $\{\psi_k(t)\}_{k=1}^K$. It is reasonable to expect that $w(t)$ is moderately smooth. However, the raw PCs of the covariance matrix of $\bX$ are potentially rough. To obtain a smooth orthonormal basis, we use fast covariance estimation (FACE) proposed by \cite{Xiao2014a} to obtain the eigenfunctions of a smoothed covariance matrix, as implemented in the {\tt R} package {\tt refund} \citep{Crainiceanu2014}. There are several potential alternative approaches. In Appendix C of the Supplementary Materials we consider pre-smoothing the exposures and using the PCs of the covariance of the smoothed data. \cite{Morris2015a} discussed several methods for regularizing functional predictors that are sampled sparsely or on irregular grids.

The orthonormal PC basis facilitates implementation of the constraints on $w(\cdot)$. When $w(t)=\sum_{k=1}^{K}\theta_k\psi_k(t)$, the constraint $\int_{t\in\mathcal{T}}  w(t)^2dt=1$ is satisfied if and only if $\|\btheta\|=1$, where $\btheta=(\theta_1,\dots,\theta_k)^T$. Additionally, $\int_{t\in\mathcal{T}}  w(t)dt\ge0$ holds for a set of observed times if and only if $\mathbf{1}^T\bPsi\btheta\ge0$ where $\mathbf{1}$ is a $T$-vector of ones and $\bPsi$ is a $T\times K$ matrix with row $t$ taking the values $[\psi_1(t),\dots,\psi_K(t)]$. The resulting constrained parameter space on $\btheta$ is, therefore, defined by the surface of a unit $K$-hemiball on one side of the hyperplane defined by $\mathbf{1}^T\bPsi\btheta=0$. In some cases, this constraint can be alleviated as discussed in Section~\ref{sub:reparam}. 

Using the PC representation for both $X(t)$ and $w(t)$ the model in \eqref{eq:BDLIM} is 
\begin{equation}
g(\mu_i)=\alpha+\beta \left(\bPsi\boldsymbol\xi_i\right)^T\bPsi\btheta+ \bZ_i^T\bgamma,
\label{eq:lm1}
\end{equation}
where $\btheta=\left(\theta_1,\dots,\theta_T\right)^T$ and $\boldsymbol\xi_i=(\xi_{i1},\dots,\xi_{iK})^T$.  Because $\bPsi$ is orthonormal,  $ \left(\bPsi\boldsymbol\xi_i\right)^T\bPsi\btheta=\xi_i^T\btheta$. The  model in \eqref{eq:lm1} is easily adapted for effect heterogeneity with group specific $\beta_j$ and $\btheta_j$.

\subsection{Reparameterization to remove constraints in the linear model }\label{sub:reparam}
For the normal linear model, BDLIM-n and BDLIM-bw can be reparameterized to reduce the computational burden imposed by the constrained parameter space. For BDLIM-n, the unknown parameters $\btheta$ for the weight functions $w(t)$ are constrained to $\|\btheta\|=1$ and $\mathbf{1}^T\bPsi\btheta\ge0$; however, $\beta$ is unconstrained and $\btheta^*=\beta\btheta$ is also unconstrained in $\mathbb{R}^{K}$. Importantly, $\beta=\|\btheta^*\|\times\text{sign}(\mathbf{1}^T\bPsi\btheta^*)$, $\btheta=\btheta^*\|\btheta^*\|^{-1}$, and $\widehat{\mathbf{w}}=\bPsi\widehat\btheta$ are each uniquely identified from $\btheta^*$.  This generalizes to the BDLIM-bw with group specific $\btheta^*_j$. Hence, we reparameterize BDLIM-n and BDLIM-bw in terms of $\btheta^*$ and estimate the model with standard Markov chain Monte Carlo (MCMC) methods without any constraints on the parameters. Then the posterior sample of $\btheta^*$ can be deconvoluted into the posterior distributions of $\beta$ and $\btheta$ by partitioning each MCMC draw. This approach is not applicable for the BDLIM-b and BDLIM-w (see Appendix A of the Supplementary Materials) and we sample directly from the constrained parameter space as described in Section~\ref{sub:priors}.

\subsection{Prior specification and computation }\label{sub:priors}
We complete the model by assigning prior distributions to the unknown regression parameters. The prior for $\btheta$ is uniform over the surface of the unit $K$-hemiball for $\btheta$. The prior likelihood can be represented as proportional to a constrained mutlivariate standard normal, $\pi(\btheta)\propto\exp(-\btheta^T\btheta/2)\Ind\{\|\btheta\|=1\}\Ind\{ \mathbf{1}^T\bPsi\btheta\ge0\}$, where $\Ind\{\cdot\}$ is an indicator function.

We assign normal priors to $\beta$ and $\gamma$.  When reparameterized, $\beta$ becomes a scale parameter for $\btheta^*$. Specifically, we assume that $\beta\sim\text{N}(0,\tau^2)$ and let $\kappa=\beta^2\tau^{-2}$. Then $\btheta^*\sim\text{N}(0,\kappa\tau^2\bI)$, where $\tau$ is fixed and $\kappa\sim\chi^2_1$. The resulting model has closed form full conditionals: $\btheta^*$ can be sampled from a multivariate normal and $\kappa$  from a generalized inverse-Gaussian distribution with density function $
f(\kappa;\lambda,\chi,\psi)\propto\kappa^{\lambda-1} \exp\{-(\chi/\kappa+\psi\kappa)/2\}$, where $\lambda=-(K-1)/2$, $\psi=1$, and $\chi=\tau^{-2}\btheta^{*T}\btheta^*$. Finally, we assume a flat prior on the intercept $\alpha$ and, for the linear model with  residuals $\epsilon_i\sim\text{N}(0,\sigma^2)$, a gamma prior on the precision parameter $\sigma^{-2}$.

We use MCMC to simulate the posterior of the unknown parameters. For the BDLIM-n and BDLIM-bw in the linear model, all unknown parameters have simple conjugate forms and can be sampled  via Gibbs sampler. For BDLIM-b, BDLIM-w, and models with a non-linear link function, we propose a slice sampling approach \citep{Neal2003} based on the elliptical slice sampler proposed by \cite{Murray2009} (see Appendix B of the Supplementary Materials for details).

\subsection{Summarizing the posterior of $w(t)$ }\label{sub:post}
Summarizing the posterior distribution of $w(t)$ deserves special consideration in light of the constraint  $\int_{t\in\mathcal{T}}  w(t)^2dt=1$. It is typical to summarize the posterior distribution with the posterior mean. However, the posterior mean of $\btheta$ almost surely does not satisfy $\|\bar\btheta\|=1$ and $\int_{t\in\mathcal{T}}  w(t)^2 dt= 1$.  To obtain a point estimate for $w(t)$ such that $\int_{t\in\mathcal{T}}  w(t)^2 dt=1$ and $\int_{t\in\mathcal{T}}  w(t) dt \ge 0$, we take the posterior mean of $\btheta$ in the topology of a $K$-hemiball parameter space.  We do this by taking the Bayes estimate with respect to the loss function $L(\btheta,\widehat\btheta) = [(\btheta-\widehat\btheta)^T(\btheta-\widehat\btheta)]/\Ind\{\|\widehat\btheta\|=1\}$. The resulting estimate $\widehat\btheta$ is the posterior mean projected onto the $K$-hemiball. That is, $\widehat\btheta=\bar\btheta\|\bar\btheta\|^{-1}$ where $\bar\btheta$ is the posterior mean.  Since each draw from the posterior satisfies $\mathbf{1}^T\bPsi\btheta\ge0$, it follows that $\mathbf{1}^T\bPsi\widehat\btheta\ge0$. We identify windows of vulnerability as time periods where the pointwise 95\% posterior intervals of $\widehat{w}(t)$ do not contain 0.

A drawback to this point estimate is that in the absence of an effect ($\beta=0$) or when $\btheta$ is not well identified by the data, the posterior of $\btheta$ will reflect the prior. In this case the projected estimator can be erratic; however, the posterior interval will still reflect the uncertainty in $\widehat{w}(t)$.

The effect $\beta$ is interpretable even when a window is not identified. Because $\int_{t\in\mathcal{T}}  w(t)dt\ge0$, the effect $\beta$ and the cumulative effect $\beta\int_{t\in\mathcal{T}}  w(t)dt$ share the same significance level, i.e. $\Pr(\beta>0 | D) = \Pr(\beta\int_{t\in\mathcal{T}}  w(t)dt >0 | D)$. Hence, the $\alpha$-level posterior interval for the cumulative effect will not contain 0 if and only if the $\alpha$-level posterior interval for $\beta$ does not contain 0 and, regardless of whether a window is identified, we can conclude that there is an overall  effect.

\subsection{Comparing models and identifying the pattern of heterogeneity}\label{sub:modelcomp}
We  quantify the evidence in the data supporting each of the four potential patterns of effect heterogeneity with the mean log posterior predictive density (MLPPD). Specifically, for model $k$ (where $k=1,2,3,4$ indicates the four BDLIM variants)  $\text{MLPPD}_k = S^{-1}\sum_{s=1}^S \log\Pr\left(\mathbf{Y}|\zeta_k^{(s)}\right)$, where  $\zeta_k$ is the vector of all parameters for model $k$ and $s=1,\dots,S$ enumerates the draws from the simulated posterior distribution. We compute $\widehat{\text{MLPPD}}_k$ for each of the four  effect heterogeneity models, then normalize as 
\begin{equation}
\widehat{P}_k= \frac{\exp\left(\widehat{\text{MLPPD}}_k \right)}{\sum_{l=1}^4 \exp\left(\widehat{\text{MLPPD}}_l \right)},
\label{eq:modelprob}
\end{equation}
In the simulation study presented in Section~\ref{sub:simB}, we compare the performance of computed using the normalized MLPPD  to identify the correct model to that of the deviance information criterion \citep[DIC,][]{Spiegelhalter2002}. Note that DIC is $-2\text{MLPPD}+p_D$, where $p_D$ is an additional penalty for model size. We show that normalized MLPPD is less likely to identify a misspecified pattern of effect heterogeneity.

\section{Simulation}\label{s:sim}
\subsection{Simulation overview}\label{s:simoverview}
We tested the performance of BDLIM with two simulation studies. Simulation A compares the BDLIM-n to a DLM when interest focuses on estimation of  the effect of a time-varying exposure in a single group. Results suggest that the two methods perform similarly. We have relegated most of the details for this simulation to Appendix C of the Supplementary Materials. We also compared tuning choices for BDLIM, including  using different numbers of PCs and using natural splines to pre-smooth the exposures instead of smoothing the covariance matrix of $\bX$ with FACE, in Appendix C of the Supplementary Materials. Simulation B highlights the advantage of BDLIM  for subgroup analyses and effect heterogeneity estimation.

For both simulations we used the observed exposures and covariates from the birth weight analysis of the ACCESS data. Using observed weekly air pollution levels during pregnancies ensures that the exposure data have realistic temporal trends and autocorrelations. Hence, the data contain $n=506$ individuals with exposures measured at 37 evenly spaced time points, ten binary covariates, three continuous covariates, and an intercept. For the second simulation we divided the data into two groups, 239 girls ($j=0$) and 267 boys ($j=1$). 

For each scenario we simulated 1000 datasets and analyzed them with BDLIM and DLM.  For BDLIM we used 15 knots to estimate the covariance matrix. We used N$(0,10^2)$ priors on $\beta$ and $\gamma$ and used the first $K$ PCs that explain 99\% of the variation in exposure. We assumed Gaussian errors and put a gamma$(0.01,0.01)$ prior on $\sigma^{-2}$. For DLM we used natural cubic splines with flat priors on the regression coefficients. We fit the DLM with degrees of freedom ranging from three to ten and have presented results only from the best performing model.

\subsection{Simulation A: Comparison to DLM with no effect heterogeneity}\label{sub:simA}
This simulation compares BDLIM-n to DLM when there is no effect heterogeneity. Both models are correctly specified but use different parameterizations and basis functions.

\begin{figure}[!p]
\centering
 \includegraphics[trim=0mm 0mm 0mm 0mm, width=1\textwidth]{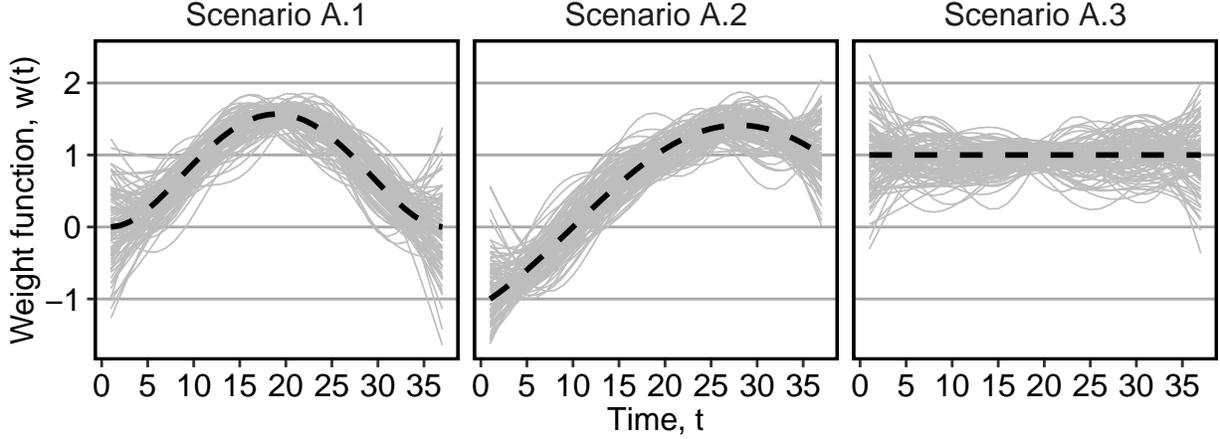}
\caption{Estimated weight functions $\widehat{w}(t)$ for simulation A. The grey lines show the estimated weight functions from BDLIM-n for the first 100 datasets. The thick black dashed line is the true weight functions.}
\label{fig:estweights}
\end{figure}

We used three scenarios each with data simulated from a different weight function:
\begin{eqnarray}
w^1(t) &=& \sqrt{\frac{t^4(1-t)^{4}}{\mathcal{B}(5,5)}} \label{eq:wt1}\\
w^2(t) &=& \frac{\sin(t\pi - \pi/4)}{\sqrt{(T^{-1}\sum_{k\in 1}^{T}\sin(t_k\pi - \pi/4)^2)}}\\
w^3(t) &=& 1,
\end{eqnarray}
where $\mathcal{B}(\cdot,\cdot)$ is a beta function and $t$ is scaled to the unit interval. The superscripts identify the weight functions and correspond to the three scenarios in simulation A.  The intercept and regression coefficients for the covariates are simulated as standard normals and we generated independent normal residuals with zero mean and standard deviation of six.   Figure~\ref{fig:estweights} shows the weight functions and the first 100 estimated weight functions using BDLIM-n.

Both BDLIM-n and DLM can estimate the total time-varying effect $\beta w(t)$, while only BDLIM-n individually identifies $\beta$ and $w(t)$. For this reason we focused the comparison on estimation of $\beta w(t)$ and the cumulative effect $37^{-1}\sum_{t=1}^{37}\beta w(t)$. Supplemental Table 1 shows that the models performed similarly.  The models had similar model fit as measured by DIC, had similar RMSE and coverage near 95\% for the estimate of $\beta w(t)$. For the cumulative effect, $T^{-1}\sum_{t=1}^T\beta w(t)$, we found that the bias and RMSE were similar for the methods and that both  had posterior interval coverage near 95\%. Hence, when there is no subgroup-specific analysis there is no information lost by using BDLIM instead of DLM.

\subsection{Simulation B: Performance with effect heterogeneity}\label{sub:simB}
The second simulation scenario compares BDLIM using the four parameterizations for exposure effect heterogeneity. The five simulation scenarios are as follows:
\begin{enumerate}
\item B.1: $w_1(t)=w^1(t)$, $\beta_1=0.1$, $w_2(t)=w^1(t)$, $\beta_2=0.1$, no heterogeneity.
\item B.2 : $w_1(t)=w^1(t)$, $\beta_1=0.1$, $w_2(t)=w^1(t)$, $\beta_2=-0.2$, heterogeneity in  $\beta$ only.
\item B.3: $w_1(t)=w^1(t)$, $\beta_1=0.1$, $w_2(t)=w^1(t)$, $\beta_2=0.0$, one group with no effect.
\item B.4: $w_1(t)=w^1(t)$, $\beta_1=0.1$, $w_2(t)=w^2(t)$, $\beta_2=0.2$, heterogeneity in $w(t)$ and $\beta$.
\item B.5: $w_1(t)=w^1(t)$, $\beta_1=0.1$, $w_2(t)=w^2(t)$, $\beta_2=0.1$,  heterogeneity in $w(t)$ only.
\end{enumerate}

For each simulation scenario, Table~\ref{tab:sim2dic} shows the model fits from each model using normalized MLPPD,  as described in Section~\ref{sub:modelcomp},  and DIC. For each of the five scenarios, there is more than one correctly specified model because  BDLIM-bw is correctly specified under any of the four forms of effect heterogeneity. Use of MLPPD to identify the best fitting model (indicated with a $*$ in Table~\ref{tab:sim2dic}) selects one of the correctly specified models with very high probability (at least 95\% for all scenarios) and almost never selects a misspecified model. In contrast, DIC selects the simplest, correctly specified model (shown in bold in Table~\ref{tab:sim2dic}) at a higher rate but also identifies a misspecified model at a higher rate for scenarios B.4 and B.5. Therefore, we conclude that use of MLPPD is a more conservative choice in that it selects a misspecified model at a lower rate at the cost of less power to rule out the most general BDLIM-bw model.
\begin{table}[!p]
\centering
\caption{Comparison of model fit with four BDLIM parameterizations for simulation B. The top panel show the mean MLPPD across the 1000 simulated datasets. The second panel shows the proportion of times each parameterization ranked as the best fitting model based on MLPPD. The third panel shows the average DIC for each parameterization. The fourth panel shows the proportion of times each model was selected as the best fitting model based on DIC. Numbers in bold indicate that that model is the simplest model that is correctly specified while an asterisk ($^*$) indicates that the model is correctly specified. Note that multiple models can be correctly specified and the BDLIM-bw is always correctly specified.} 
\label{tab:sim2dic}
\begin{tabular}{lccccc}
  \hline
  & \multicolumn{5}{c}{Scenario} \\ \cline{2-6}
   &  B.1& B.2&  B.3 & B.4 &  B.5 \\ \hline
 \multicolumn{6}{l}{Mean log posterior predictive distribution} \\
BDLIM-b &    0.30$^*$ & \textbf{   0.61}$^*$ & \textbf{   0.64}$^*$ &    0.02 &    0.06  \\ 
BDLIM-bw &    0.21$^*$ &    0.39$^*$ &    0.32$^*$ & \textbf{   0.97}$^*$$^*$ &    0.42$^*$  \\ 
BDLIM-n & \textbf{   0.24}$^*$ &    0.00 &    0.00 &    0.00 &    0.02  \\ 
BDLIM-w &    0.24$^*$ &    0.00 &    0.04 &    0.00 & \textbf{   0.50}$^*$  \\ 
  \hline
   \multicolumn{6}{l}{Proportion model selected  using mean log posterior predictive distribution} \\
BDLIM-b &    0.26$^*$ & \textbf{   0.63}$^*$ & \textbf{   0.68}$^*$ &    0.02 &    0.05  \\ 
BDLIM-bw &    0.13$^*$ &    0.37$^*$ &    0.28$^*$ & \textbf{   0.98}$^*$$^*$ &    0.28$^*$  \\ 
BDLIM-n & \textbf{   0.30}$^*$ &    0.00 &    0.00 &    0.00 &    0.02  \\ 
BDLIM-w &    0.31$^*$ &    0.00 &    0.04 &    0.00 & \textbf{   0.65}$^*$  \\ 
   \hline
   \multicolumn{6}{l}{Mean DIC} \\
BDLIM-b & 3270.32$^*$ & \textbf{3270.85}$^*$ & \textbf{3269.77}$^*$ & 3287.60 & 3279.42  \\ 
BDLIM-bw & 3274.57$^*$ & 3275.51$^*$ & 3273.16$^*$ & \textbf{3276.02}$^*$$^*$ & 3274.93$^*$  \\ 
BDLIM-n & \textbf{3270.44}$^*$ & 3504.49 & 3302.76 & 3300.47 & 3281.10  \\ 
BDLIM-w & 3273.94$^*$ & 3360.48 & 3283.67 & 3293.79 & \textbf{3274.31}$^*$  \\ 
   \hline
   \multicolumn{6}{l}{Proportion model selected using DIC} \\
BDLIM-b &    0.31$^*$ & \textbf{   0.92}$^*$ & \textbf{   0.85}$^*$ &    0.11 &    0.15  \\ 
BDLIM-bw &    0.02$^*$ &    0.08$^*$ &    0.14$^*$ & \textbf{   0.89}$^*$$^*$ &    0.12$^*$  \\ 
BDLIM-n & \textbf{   0.56}$^*$ &    0.00 &    0.00 &    0.00 &    0.10  \\ 
BDLIM-w &    0.11$^*$ &    0.00 &    0.02 &    0.00 & \textbf{   0.63}$^*$  \\ 
  \hline
\end{tabular}
\end{table}

Table~\ref{tab:sim2inf} summarizes inference for $\beta$ and $w(t)$ for the different BDLIM models. Overall, using the model that matches the pattern of heterogeneity (indicated by bold in Table~\ref{tab:sim2inf}) in the data provides the best inference. This is most notable for scenario B.4 where both $\beta_j$ and $w_j(t)$ vary by group and BDLIM-bw provides accurate inference while the other approaches have greater bias, larger RMSE, and lower coverage for both $\beta$ and $w(t)$. Similarly, for scenarios B.2 and B.3, BDLIM-b results in improved estimation of $w(t)$ by sharing information across groups. To a lesser extent, BDLIM-w yields $\beta$ estimates with lower RMSE in scenario B.5. 

In summary, BDLIM-n is nearly identical to a standard distributed lag model when there is no heterogeneity across groups. When there is heterogeneity across groups, using BLDIM we can identify a correctly specified model with high probability. When that model is a reduced model, BDLIM results in improved inference of the weight function and effect size.
\begin{table}[!p]
\centering
\small
\caption{Simulation results for the inference on $\beta$ and $w(t)$ using BDLIM for simulation B. For each scenario 1000 datasets were fit. The table shows the bias, RMSE, and 95\% credible interval coverage for $\widehat\beta$ and the RMSE and 95\% credible interval coverage for $\widehat{w}(t)$. The bold rows indicate the model that was most frequently selected as the MLPPD model for that simulation scenario.} 
\label{tab:sim2inf}
\begin{tabular}{llccccc}
  \hline
   &  & \multicolumn{3}{c}{Inference for $\beta$}& \multicolumn{2}{c}{Inference for $w(t)$}\\   \cmidrule(lr){3-5}\cmidrule(lr){6-7}  
      & Group & Bias & RMSE & Cover & RMSE & Cover\\  
      \hline\multicolumn{7}{l}{Scenario B.1: No heterogeneity}\\
BDLIM-b & Female &  0.000 & 0.011 & 0.973 & 0.248 & 0.954  \\ 
BDLIM-b & Male & -0.001 & 0.012 & 0.970 & 0.248 & 0.954  \\ 
BDLIM-bw & Female &  0.000 & 0.011 & 0.976 & 0.321 & 0.958  \\ 
BDLIM-bw & Male & -0.001 & 0.012 & 0.980 & 0.317 & 0.959  \\ 
\textbf{BDLIM-n} & Female &  0.005 & 0.011 & 0.947 & 0.258 & 0.947  \\ 
\textbf{BDLIM-n} & Male &  0.005 & 0.011 & 0.947 & 0.258 & 0.947  \\ 
BDLIM-w & Female & -0.001 & 0.010 & 0.970 & 0.317 & 0.957  \\ 
BDLIM-w & Male & -0.001 & 0.010 & 0.970 & 0.316 & 0.958  \\ 
          \hline\multicolumn{7}{l}{Scenario B.2: Heterogeneity in  $\beta$ only}\\  
\textbf{BDLIM-b} & Female &  0.000 & 0.012 & 0.975 & 0.166 & 0.947  \\ 
\textbf{BDLIM-b} & Male &  0.001 & 0.017 & 0.954 & 0.166 & 0.947  \\ 
BDLIM-bw & Female &  0.008 & 0.013 & 0.964 & 0.335 & 0.944  \\ 
BDLIM-bw & Male &  0.007 & 0.013 & 0.934 & 0.173 & 0.956  \\ 
BDLIM-n & Female & -0.176 & 0.176 & 0.000 & 0.424 & 0.964  \\ 
BDLIM-n & Male &  0.124 & 0.125 & 0.000 & 0.424 & 0.964  \\ 
BDLIM-w & Female & -0.285 & 0.285 & 0.000 & 1.465 & 0.167  \\ 
BDLIM-w & Male &  0.015 & 0.018 & 0.806 & 0.163 & 0.966  \\ 
        \hline\multicolumn{7}{l}{Scenario B.3: One group (males) with no effect}\\
\textbf{BDLIM-b} & Female &  0.002 & 0.012 & 0.968 & 0.318 & 0.950  \\ 
\textbf{BDLIM-b} & Male & -0.002 & 0.012 & 0.957 & 0.318 & 0.950  \\ 
BDLIM-bw & Female & -0.010 & 0.014 & 0.926 & 0.280 & 0.960  \\ 
BDLIM-bw & Male & -0.006 & 0.020 & 0.970 & 0.650 & 0.983  \\ 
BDLIM-n & Female & -0.041 & 0.042 & 0.037 & 0.512 & 0.927  \\ 
BDLIM-n & Male &  0.059 & 0.059 & 0.015 & 0.512 & 0.927  \\ 
BDLIM-w & Female & -0.022 & 0.024 & 0.538 & 0.283 & 0.980  \\ 
BDLIM-w & Male &  0.078 & 0.079 & 0.000 & 1.307 & 0.527  \\ 
        \hline\multicolumn{7}{l}{Scenario B.4: Heterogeneity in both $w(t)$ and $\beta$ }\\
BDLIM-b & Female &  0.006 & 0.018 & 0.941 & 0.648 & 0.151  \\ 
BDLIM-b & Male & -0.022 & 0.042 & 0.793 & 0.307 & 0.794  \\ 
\textbf{BDLIM-bw} & Female &  0.003 & 0.012 & 0.970 & 0.352 & 0.943  \\ 
\textbf{BDLIM-bw} & Male & -0.010 & 0.021 & 0.919 & 0.172 & 0.943  \\ 
BDLIM-n & Female &  0.048 & 0.049 & 0.031 & 0.559 & 0.236  \\ 
BDLIM-n & Male & -0.052 & 0.054 & 0.037 & 0.318 & 0.658  \\ 
BDLIM-w & Female &  0.043 & 0.045 & 0.039 & 0.561 & 0.756  \\ 
BDLIM-w & Male & -0.057 & 0.058 & 0.008 & 0.281 & 0.849  \\ 
           \hline\multicolumn{7}{l}{Scenario B.5: Heterogeneity in $w(t)$ only}\\
BDLIM-b & Female &  0.003 & 0.013 & 0.974 & 0.399 & 0.795  \\ 
BDLIM-b & Male & -0.022 & 0.028 & 0.674 & 0.580 & 0.537  \\ 
BDLIM-bw & Female &  0.000 & 0.011 & 0.976 & 0.315 & 0.960  \\ 
BDLIM-bw & Male & -0.007 & 0.018 & 0.944 & 0.355 & 0.949  \\ 
BDLIM-n & Female & -0.002 & 0.011 & 0.968 & 0.472 & 0.667  \\ 
BDLIM-n & Male & -0.002 & 0.011 & 0.968 & 0.490 & 0.655  \\ 
\textbf{BDLIM-w} & Female & -0.003 & 0.011 & 0.970 & 0.312 & 0.962  \\ 
\textbf{BDLIM-w} & Male & -0.003 & 0.011 & 0.970 & 0.328 & 0.950  \\ 

       \hline
\end{tabular}
\end{table}

\section{Analysis of prenatal air pollution exposure}\label{s:da}
\subsection{Impact of sex and maternal obesity on air pollution effects on birth weight}\label{sub:bwz}
We used BDLIM to estimate the association between PM$_{2.5}$ and BWGA $z$-score. Following the analysis of \cite{Lakshmanan2015}, we estimated this association by child sex and maternal obesity status. Of the 506 children with complete data including BWGA $z$-score, there were 155 females with non-obese mothers, 182 males with non-obese mothers,  84 females with obese mothers, and 85 males with obese mothers. \cite{Lakshmanan2015} associated average PM$_{2.5}$ over the entire pregnancy with  BWGA $z$-score. Here, we estimate the association using the four variations of BDLIM to identify windows of vulnerability and use MLPPD to select the best fitting model. We assumed a normal linear model and used the same priors as described for the simulation in Section~\ref{s:simoverview} and set $K$ to explain 99\% of the variance in $X(t)$.

The BDLIM-b model had the highest normalized MLPPD at 0.96. The other models were 0.02 for BDLIM-bw, 0.01 for BDLIM-w, and 0.01 for BDLIM-n. Hence, we present results from the BDLIM-b, which assumes a single weight functions $w(t)$ shared by all four groups but group specific $\beta_j$.

\begin{figure}[!p]
\centering
\subfloat[$\widehat\beta_j$]{
 \includegraphics[trim=0mm 0mm 0mm 0mm, width=.25\textwidth]{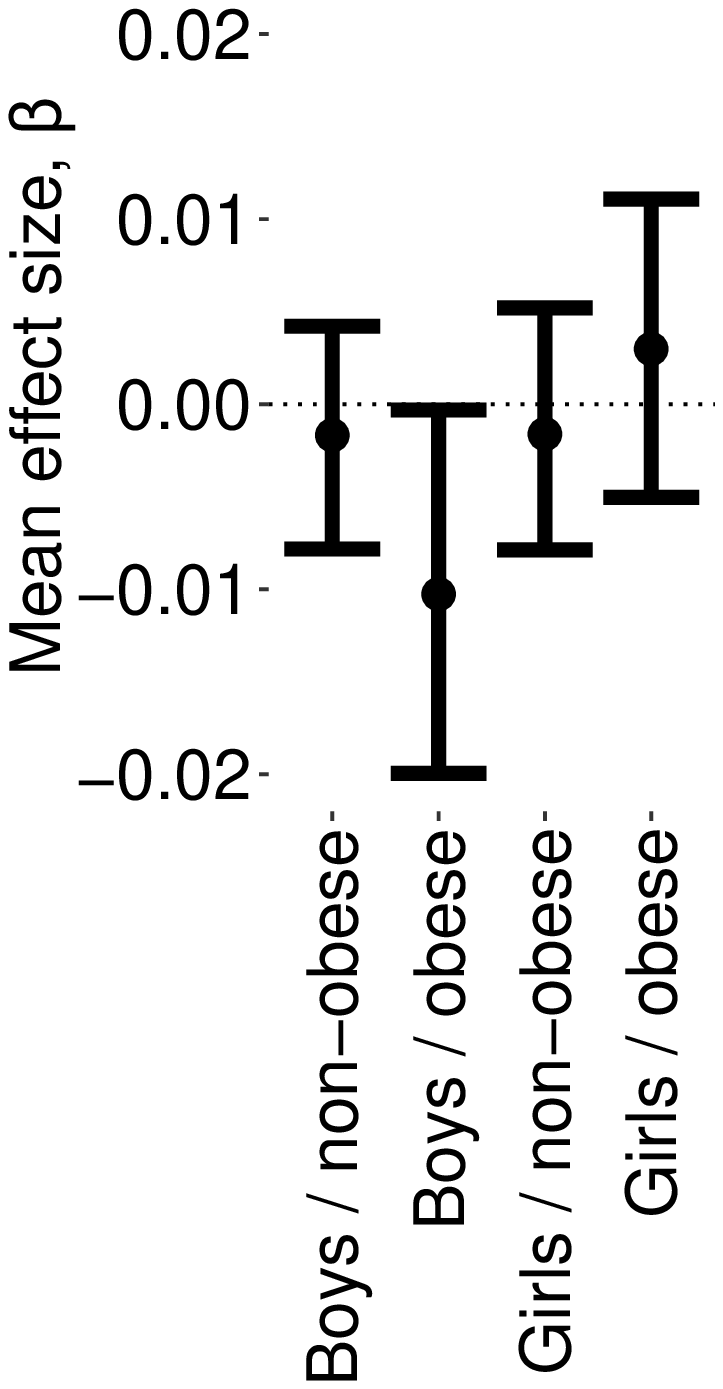} 
 \label{fig:est1bw_b}}
\subfloat[$\widehat{w}(t)$]{
 \includegraphics[trim=0mm 0mm 0mm 0mm, width=.75\textwidth]{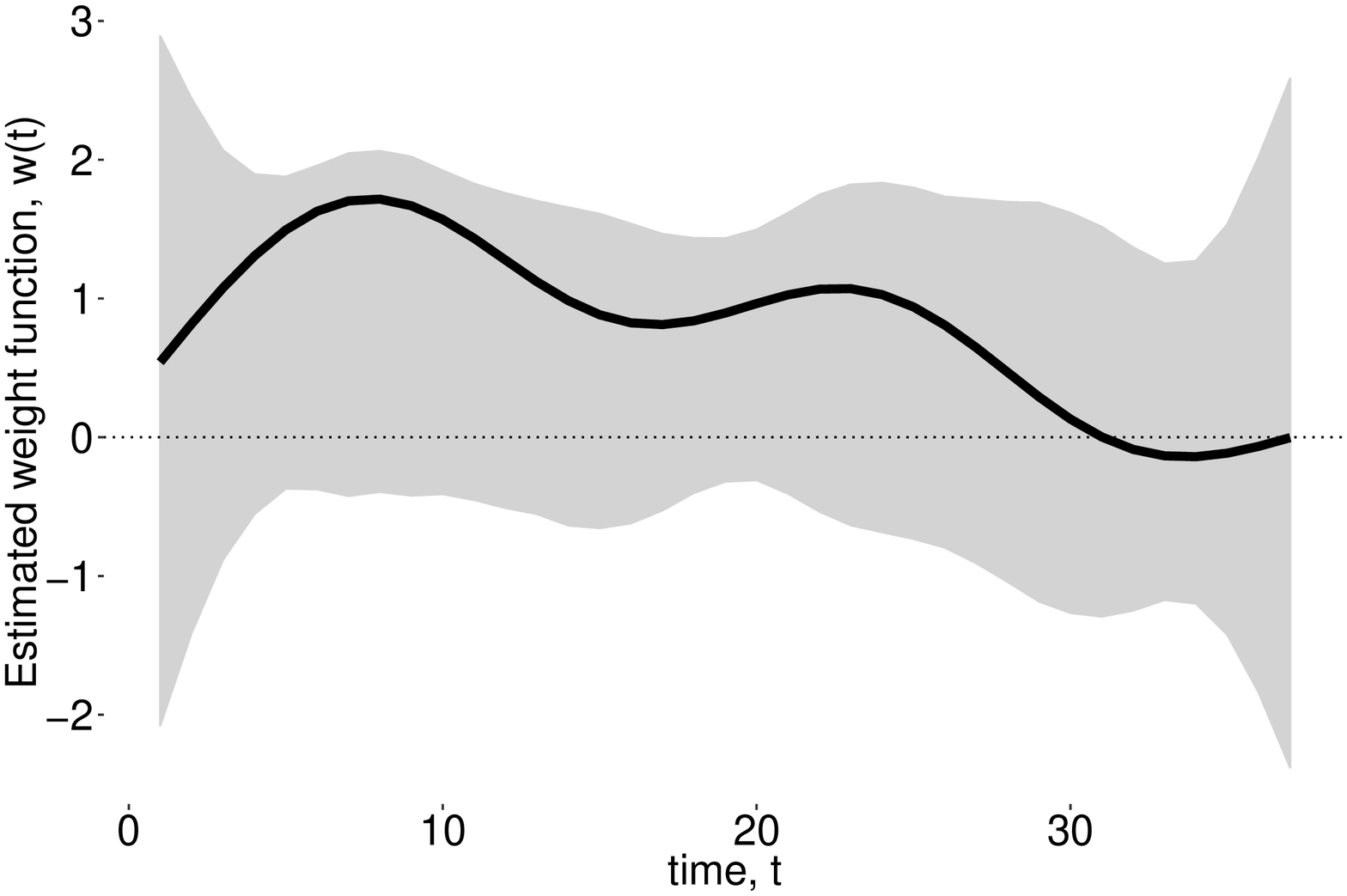}
 \label{fig:est1bw_w}
 }
\caption{Estimated group specific effect sizes $\widehat\beta_j$ (left panel) and weight function $\widehat{w}(t)$ (right panel) using the BDLIM-b model for the BWGA $z$-score analysis.}
\label{fig:est1bw}
\end{figure}

Figure~\ref{fig:est1bw_b} shows the estimated group-specific effects $\widehat{\beta}_j$. The results are consistent with those reported by \cite{Lakshmanan2015}, with a negative association between PM$_{2.5}$ and BWGA $z$-score among boys with obese mothers but not in the other groups. In addition, the posterior probability of a pairwise difference between boys with obese mothers and the other three groups range from 0.93 to 0.97. The posterior difference for the pairwise comparisons between the three non-significant groups range from 0.50 to 0.84 suggesting little evidence of differences between those groups. Further, because the model is saturated we can perform an ANOVA decomposition on the posterior sample to investigate main effects of sex and obesity as well as an interaction effect (see Appendix D of the Supplementary Materials for details).

The estimated weight function $\widehat{w}(t)$ (Figure~\ref{fig:est1bw_w}) shows a trend of increased vulnerability in the earlier part of pregnancy, approximately weeks 5 through 20. Although we do not identify a window with high probability, the estimated cumulative effect over the full pregnancy always has the same sign and significance level as $\hat\beta_j$. For males infants with obese mothers we estimate a cumulative effect of -0.225 with 95\% credible interval (-0.476, -0.001). Hence, the results are suggestive of a negative association between PM$_{2.5}$ exposures in early pregnancy and lower BWGA $z$-score among boys with obese mothers.

The estimated time-varying effects $\beta_j w(t)$ are presented in Appendix D of the Supplementary Materials. Results from the BDLIM-bw model are presented there as well. Comparing the results using BDLIM-bw and BDLIM-b shows that the model with a shared weight function (BDLIM-b) is more suggestive of the location of the window of vulnerability and has, on average, 14\% smaller posterior standard deviation for $\widehat\beta_j$.   Therefore, there is evidence that the magnitude of the effect varies across groups, but no evidence that the timing of the window varies across groups primarily because there is no effect in three of the four groups. 

\subsection{Sex-specific effects of prenatal air pollution on asthma incidence}\label{sub:asthma}
We next used a logistic  BDLIM to estimate sex-specific associations between PM$_{2.5}$ and childhood asthma incidence in the ACCESS  cohort. \cite{Hsu2015} analyzed these data by stratifying by sex and applying a standard DLM to data in each stratum. Here we assess whether the magnitude of the effect, the timing of the effect, or both vary by sex.  This analysis included data from 544 births with complete data including asthma. Again, BDLIM-b was the best fitting model for the asthma analysis with normalized MLPPD of 0.43. The other were 0.21 for BDLIM-bw, 0.20 for BDLIM-n, and 0.17 for BDLIM-w. We describe the results from the BDLIM-b model. 

\begin{figure}[!p]
\centering
\subfloat[$\widehat\beta_j$]{
 \includegraphics[trim=0mm 0mm 0mm 0mm, width=.25\textwidth]{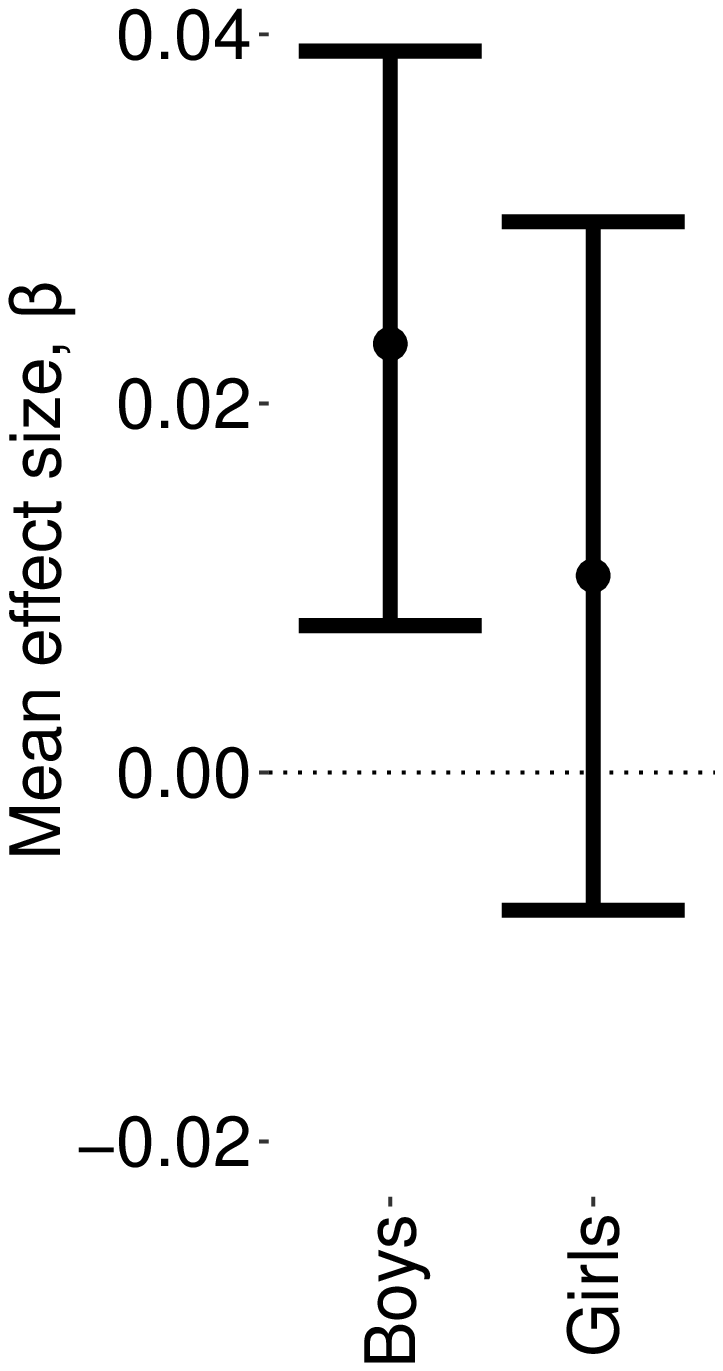}
 \label{fig:est1asthma_b}}
\subfloat[$\widehat{w}(t)$]{
 \includegraphics[trim=0mm 0mm 0mm 0mm, width=.75\textwidth]{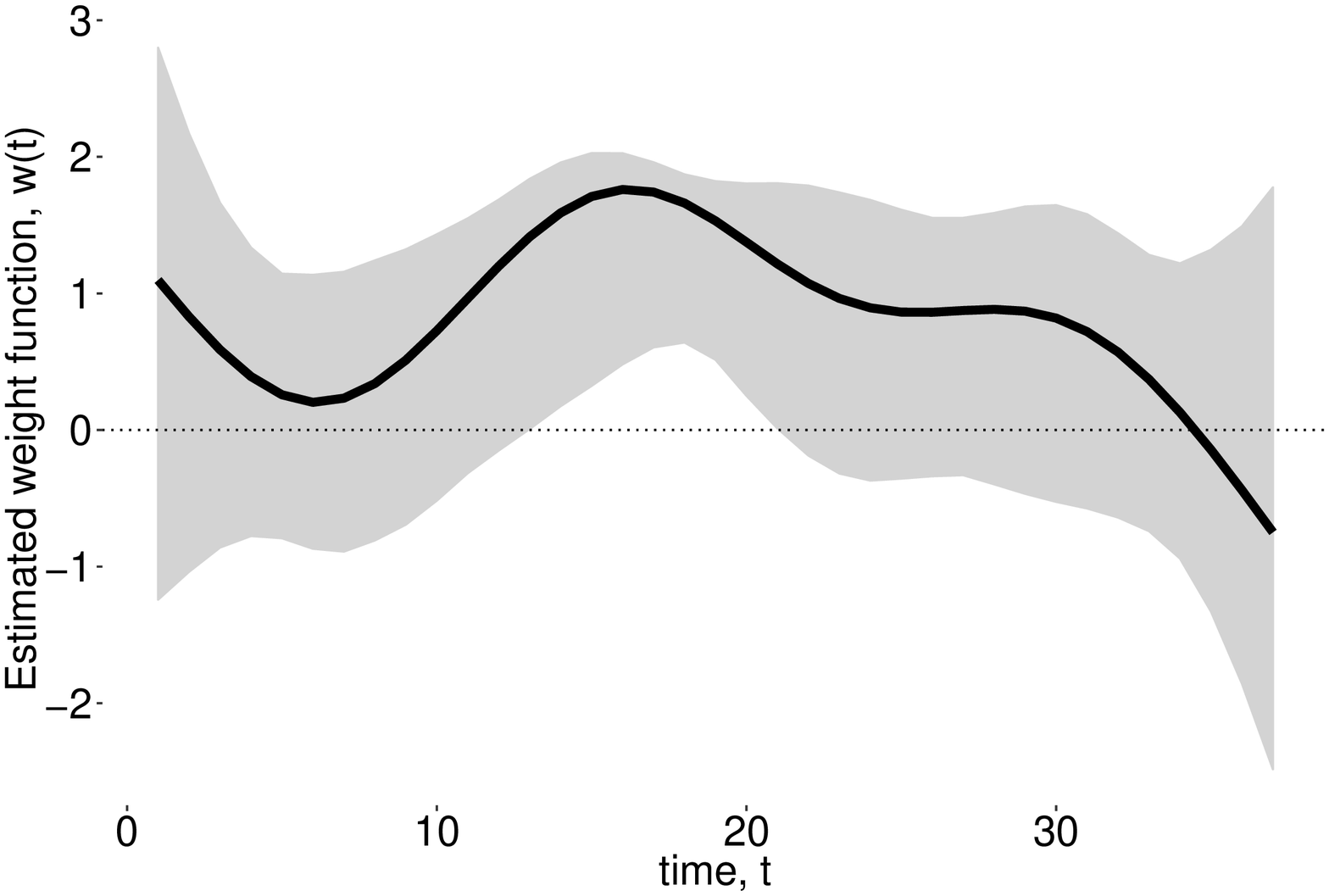}
 \label{fig:est1asthma_w}}
 
\caption{Estimated group specific effect sizes $\widehat\beta_j$ (left panel) and weight function $\widehat{w}(t)$ (right panel) using the BDLIM-b model on the asthma analysis.}
\label{fig:est1asthma}
\end{figure}

Figure~\ref{fig:est1asthma} shows the estimated weight function $\widehat{w}(t)$ and the estimated group-specific effects $\widehat\beta_j$ using BDLIM-b. Figure~\ref{fig:est1asthma_b} shows a positive and statistically significant association between PM$_{2.5}$ exposure and asthma incidence in  boys but not in girls. The weight function  (Figure~\ref{fig:est1asthma_w}) identifies a window of vulnerability in weeks 13-21.  Hence,  PM$_{2.5}$ exposures during the weeks 13-21 were positively associated with asthma among boys. This is comparable to the widows identified by \cite{Hsu2015} which were 12-26 weeks for boys using a sex-stratified analysis and 14-20 weeks when testing sex differences. Web Figure 5 shows the estimated sex-specific odds ratios for  a 10$\mu g$/m$^3$ increase in PM$_{2.5}$, which peak around 1.25 for boys. The results using BDLIM-bw were very similar and are included in Appendix E of the Supplementary Materials; however, BDLIM-b yielded posterior standard deviations of $\widehat\beta_j$ that were 10\% smaller than their BDLIM-bw counterparts.

\section{Discussion}
In this paper we have proposed BDLIM as a new tool that can be used  estimate effect heterogeneity in time-varying exposures. This addresses a critical methodological gap for simultaneous estimation of windows of vulnerability and identifying susceptible subpopulations. Specifically, BDLIM allows for estimation of effect heterogeneity when subgroups have a common window of vulnerability but different effects within the window (BDLIM-b) or when subgroups have the same effect size but in different windows (BDLIM-w). In these scenarios, the resulting estimates had reduced bias and RMSE,  an advantage of pooling information across groups when the effects are not different.  We demonstrated this advantage both in the simulation study and the data analysis. 

The proposed approach partitions the time-varying effect into two components. The first is a constrained functional predictor that captures the temporal variation in the effect and  identifies windows of vulnerability. The second component is a scalar effect size that quantifies the effect within the window.  In some situations the constraint on the parameters of the weight function can be removed by reparameterizing the model. In this case the scalar effect size becomes a scale parameter in the model with a generalized inverse-Gaussian full conditional, which allows for simulating the posterior with a Gibbs sampler. In other situations, including generalized linear models, we use a slice sampler to efficiently simulate the posterior of the contained parameters. The constraints of the weight function make summarizing the posterior of the weight function difficult because the posterior mean does not satisfy the constraints. We address this by using a point estimate that is the Bayes estimate with respect to a non-standard loss function specifically chosen to yield a posterior summary that satisfies the constraints. We then identify windows of vulnerability where the pointwise posterior interval does not contain zero. 

We analyzed data from the ACCESS cohort on the association between prenatal exposure to PM$_{2.5}$ and both birth weight and asthma incidence.  In both cases the BDLIM analyses suggested  a common window of vulnerability but different effect sizes within the window for both outcomes. Hence, in both analyses the model providing the best fit to the data  could not have been estimated using existing methods. Our results identified a window of weeks 13-21 where PM$_{2.5}$ exposures were associated with increased asthma incidence in boys but not in girls. In the  analysis of the birth weight data, there was strong evidence of a negative association between PM$_{2.5}$ in the earlier part of pregnancy and decreased BWGA $z$-score among boys born to obese mothers.

The proposed approach assumes that $\beta$ and/or $w(t)$ are the same for all groups or different for all groups. When there are more than two groups it may be of interest to understand if only certain pairs of groups share one or more component. In the BWGA $z$-score analysis we performed an \emph{a posteriori} ANOVA decomposition to investigate if the main effect of child sex or maternal obesity status is important. This could be done because the model for $\beta$ was saturated. Another simple approach would be to define groups based on preliminary analysis and rerun the model. In the BWGA $z$-score analysis this would mean two groups: boys with obese mothers in one group and everyone else in another. A more sophisticated extension would be to extend the approach to directly model how covariates influence each component, such as modeling the weight functions as $w_j(t) = w_0(t) + w_{sex}(t) + w_{obese}(t)$.

Identifying susceptible populations and windows of vulnerability are critical areas of future research as highlighted in the NIEHS strategic plan \citep{NIEHS2012}. BDLIM provides an essential tool for  simultaneous identification of  susceptible populations and critical windows of vulnerability when estimating of  the health effects of environmental exposures.

\section{Supplementary Material}
Supplementary Material is available upon request.

\section*{Funding}
The ACCESS study has been supported by grants R01 ES010932, R01 ES013744; U01 HL072494, and R01 HL080674 (Wright RJ, PI). This work was supported by USEPA grant 834798 and NIH grants (ES020871, ES007142, CA134294, ES000002, P30 ES023515). This publication's contents are solely the responsibility of the grantee and do not necessarily represent the official views of the US EPA.

\end{document}